\documentclass[twocolumn,showpacs,preprintnumbers]{revtex4}
\usepackage{graphicx}
\usepackage{dcolumn}     
\usepackage{bm}          
\usepackage{subfigure}
\usepackage{amsmath}
\usepackage{amssymb}
\usepackage{setspace}
\usepackage{array}
\usepackage{latexsym}
\newcommand{\ds}{\displaystyle}
\newcommand{\scs}{\scriptscriptstyle}

\newcommand{\pr}{\scriptscriptstyle \prime}

\def\d{\delta} 
 
\def\g{\gamma}

\def\a{\alpha} 
\def\o{\omega}

\def\e{\epsilon}

\def\DLE{{\cal D}} 
\begin{document}
\title{Cooling non--linear lattices toward energy localisation} 
\author{Francesco Piazza}
\affiliation{I.N.F.M.
UdR di Firenze, Via G. Sansone 1, 50019 Sesto F.no, ITALY}
\author{Stefano Lepri}
\affiliation{Dipartimento di Energetica ``S. Stecco'', Via S. Marta 3,  
50139 Firenze and I.N.F.M. UdR di
Firenze, Italy}
\author{Roberto Livi}
\affiliation{Dipartimento di Fisica and I.N.F.M. UdR di
Firenze\\ Via G. Sansone 1, 50019 Sesto F.no, ITALY}
\begin{abstract}
We describe the energy relaxation process produced by surface damping on lattices of
classical anharmonic oscillators. Spontaneous emergence of localised vibrations
dramatically slows down dissipation and gives rise to quasi-stationary states where energy
is trapped in the form of a gas of weakly interacting discrete breathers. In one dimension
(1D), strong enough on--site coupling may yield stretched--exponential relaxation which is
reminiscent of glassy dynamics. We illustrate the mechanism generating localised
structures and discuss the crucial role of the boundary conditions. For two--dimensional
(2D) lattices, the existence of a gap in the  breather spectrum causes the localisation
process to become activated.  A statistical analysis of the resulting quasi-stationary
state through the distribution of breathers' energies yield information on their effective
interactions. 
\end{abstract}
\maketitle
{\bf Discrete breathers have been widely analysed as mathematical objects, but the relevance 
of their role in physical systems is still debated. In fact, any mechanism yielding the spontaneous
formation of  breathers is expected to provide interesting indications  in this direction.
In this respect, the phenomenon of  relaxation to breather states by cooling lattices from
their boundaries is one of the most interesting examples. Here we provide a comprehensive
description of this phenomenon in low dimensional lattices. In particular, we review former
results and add new information, mainly  concerning the 2D case. This study is based on a 
combination of numerics and theoretical arguments, while we analyse both dynamical and statistical
features of the problem to clarify the whole scenario.}
\bigskip
{
\section{Introduction}

Nonlinearity has revealed one of the key ingredients for describing many relevant features of
different states of matter. In the realm of lattice dynamics,  scattering processes among
phonons, propagation of solitary waves and slow energy relaxation are typical examples.
Recently, considerable efforts have been devoted to the study of periodic, localised,
non--linear lattice excitations, named "breathers"  \cite{breath1} . They are quite peculiar
(but generic) objects emerging from the interplay of nonlinearity and space discreteness
\cite{breath2}. Their mathematical properties, like existence, stability, mobility etc. have
been progressively unveiled, while they have been identified in many different scenarios. 

Their role in non--equilibrium dynamics seems to be particularly fascinating. An example 
is the relaxation to energy equipartition of short--wavelength fluctuations~\cite{chao_breath}. 
Another interesting scenario where breathers are found to emerge spontaneously is observed upon cooling
the lattice at its boundaries \cite{Aubry1,Aubry2,noi}, i.e. by considering a non--equilibrium process
in which energy exchange with the environment is much faster at the surface than in the bulk. The
numerical simulations neatly show that the energy dissipation rate may be significantly affected by the
spontaneous excitation of breathers. As the latter exhibit a very weak interaction among themselves
and with the boundaries, the energy release undergoes a sudden slowing down and is hardly detected
on the time scales of a typical simulation. Thus, the lattice remains frozen in a
pseudo--stationary, metastable configuration which is far from thermal equilibrium, which
we shall refer to as {\em residual state}. Despite the absence of disorder, 
there is a close similarity with the glassy
behaviour observed in disordered systems, as was already pointed out \cite{Aubry1,Aubry2}.  

But what are the mechanisms leading to spontaneous localisation? As we have shown in a
previous paper~\cite{noi}, the latter is intimately related to how dissipation acts on
vibrational modes of different wavelengths. If long--wavelength phonons can be efficiently
damped out, the modulation instability of short lattice waves~\cite{Peyrard} becomes highly 
favoured and breathers can  easily emerge
from an interacting soliton--gas~\cite{chao_breath,Kosevich}.
In this respect, 
this type of non--equilibrium condition is much more effective in exciting localised modes 
than an equilibrium one (see e.g. \cite{Bishop} for a related discussion).

Spontaneous localisation upon cooling has been observed for both on--site~\cite{Aubry1,Aubry2} and
pure nearest--neighbour (n.n.)  interactions~\cite{noi,reigada1}. The two classes of models are known
to behave differently, e.g. as to  the mobility of breathers, which is much higher in the absence of
local coupling. This is also confirmed by relaxations experiments, where differences in the  energy
decay laws have been observed.  In particular, the approach to the residual state  follows a simple
exponential law in the Fermi-Pasta-Ulam (FPU) system~\cite{noi}, but turns to stretched exponential 
in the presence of local coupling~\cite{Aubry2}.

Our aim in this paper is to provide an up--to--date overview of the phenomenon of breather
localisation by cooling. Generalities about the energy release process, including models and
indicators, will be presented in Section~\ref{sec_gen}, along with the discussion of the cooling
process in an harmonic lattice and the results obtained for 1D non--linear systems. In particular we
shall survey the different cooling pathways appearing in pure n.n. and on--site potentials. A specific
subsection will be devoted to clarify the interpretation of some recent
results~\cite{reigada1,reigada2}.  In Section~\ref{sec_2D}, we present the results for the n.n. 2D
FPU  model. We want to point out that the 2D is not a straightforward extension of the 1D case.
Indeed,  the existence of an energy activation threshold for breather solutions~\cite{2D_tresh} leads
to conjecture that a thermalised state may be cooled to the residual state only above some initial
energy/temperature.  Moreover, the residual state is a static multi--breather
state, whereby in the 1D case it usually contains a single  breather.  Finally, we perform  an
analysis of the distribution of breather energies, based on a simple statistical model. 

\section{\label{sec_gen} Relaxation and localisation in 1D}

In this section we will describe the lattice models and the generalities of a typical
relaxation experiment as well as the relevant indicators used throughout the paper.  Let us consider
a chain of $N$ atoms of unit mass and denote with $u_p$ the displacement of the $p$--th particle
from its equilibrium position $pa$ ($a$ is the lattice spacing). The atoms are labelled 
by the index $p = 0, 1,\dots N-1$ and their dynamics is given by the 
equations of motion
\begin{equation}
\label{eq_mot}
\begin{split}
\ddot{u}_p =  \,  &V'(u_{p+1}-u_p) - V'(u_p-u_{p-1}) -U'(u_p) \nonumber\\
                  &-\g  \dot{u}_p \left[ \d_{p,0} +\d_{p,N-1} \right]\quad, 
\end{split}	       
\end{equation}
where $V(x)$ and $U(x)$ are the interparticle and on--site potentials, respectively. We assume
that $V'(0) = U'(0) = 0$, denoting with a prime the derivative with respect to $x$.  For
convenience, we shall adopt  non--dimensional units such that $a$ and  $V''(0)$ are set
to unity. Moreover, as we want to deal  with systems in a finite volume, we will impose either
free--end  ($u_{-1}=u_{0}$, $u_{N-1}=u_{N}$) or fixed--end  ($u_{-1} = u_{N} = 0$) boundary
conditions (BC).

The last term in eq.~(\ref{eq_mot}) represents the interaction of the atoms with a ``zero--temperature'' 
heat bath in the form of a linear damping with rate $\g$.  Following \cite{Aubry1,Aubry2}, we restricted 
to the case  in which dissipation selectively acts only on the atoms located at the chain edges. As 
mentioned above, this choice is crucial and should model a physical situation in which  energy exchange
with the environment is much faster at the surface than in the bulk. 

The general layout of a simulation  can be summarised as follows . First, an equilibrium microstate is
generated by, say, Nos\`e-Hoover (canonical) method~\cite{Aubry1}  or by letting the Hamiltonian
(microcanonical) system ($\gamma$ = 0) evolve for a sufficiently long  transient~\cite{noi}. In the
following we will follow the second strategy. The initial condition for the transient is assigned by
setting all displacements $u_p$ to zero and by drawing velocities at random from a Gaussian
distribution. The velocities are then rescaled by a suitable factor to fix the desired value of the
energy density (energy per particle) $e_0$.  The resulting set of $u_p$ and $\dot u_p$ is then used as
initial  condition to integrate  Eqs.~(\ref{eq_mot}) with $\gamma > 0$ (see again Ref.~\cite{noi} for
details).

The result of the relaxation process is the residual state, i.e. a state with a finite fraction
of the initial energy stored in  the form of breathers. Remarkably, such residual state is
characterised by a decay time scale which is several orders of magnitude longer than that of the
localisation process.

\subsection{\label{sec_harm} The harmonic lattice}
A number of features which characterise the relaxation dynamics of non--linear lattices can be better
understood by first studying the harmonic lattice, namely
\[
V(x) \; = \; \frac{1}{2} \, x^2,
\]
We anticipate that the presence of an harmonic on--site potential does not alter the main 
conclusions. We thus take $U = 0$ throughout this section. 

For small damping, an approximate analytical solution can be found by  time--dependent perturbation
theory. Let $\o^2_\a=4\sin^2({q_\a}/{2} )$ and  $\eta^\a$ ($\a = 0, 1, \dots N-1$) denote the
eigenvalues and normalized eigenvectors of the unperturbed  Hamiltonian problem,
where the allowed wave--numbers are  $q_\a = \alpha \pi/N$ and $q_\a = (\alpha+1) \pi/(N+1)$ for 
free--end and fixed--end BC, respectively.  To the lowest order in $\g$ one obtains~\cite{noi}
\begin{equation}
\label{pert_ans}
u_p(t)=\sum_{\a=0}^{N-1} \, c_\a   e^{\ds -\left( \frac{\ds 1}{\ds \tau_\a}+i \o_\a \right)t} \, \eta_p^\a.
\end{equation}
where $\eta_p^\a$ is the $p$-th component of the eigenvector $\eta^\a$ and $c_\a$ are constant
amplitudes fixed by the initial conditions. Notice that, since the number of the damped
particles is fixed, the perturbative approximation is expected to improve upon increasing the
system size $N$. The wavenumber-dependent damping rates are found to be
\begin{equation}
\label{calpha}
\frac{1}{\tau_\alpha}=
\begin{cases}
\frac{\ds 1}{\ds \tau_0} \, \cos^2 \left(\frac{\ds q_\a}{\ds 2} \right) & \text{for free--end BC} \\
\frac{\ds 1}{\ds \tau_0} \, \sin^2 (q_\a) & \text{for fixed--end BC} \quad . 
\end{cases}
\end{equation}
with $\tau_0 =  N / 2 \gamma$. The free--end and fixed--end BC systems thus show considerably
different behaviours. In the former case, the least damped modes are the short--wavelength ones
($\a \approx N$), the largest lifetime being $\tau_{\scs N-1} \approx 2 N^3 / \pi^2 \gamma$,
while the most damped modes are the ones in the vicinity of $\a = 0$, with  $\tau_0$ being the
shortest decay time.  On the contrary, for fixed ends the most damped modes are those
around the band center ($\a \approx N/2$) while short-- and long--wavelength ones
dissipate very weakly, being  $\tau_{\scs N-1} \approx N (N + 1)^2 / 2 \pi^2 \gamma$.  As we
shall illustrate in the following, such difference is crucial for the localisation in the
non--linear system.  

Using the solution~(\ref{pert_ans}), we evaluated the chain energy in the case of equipartition 
i.e. by replacing $ c^2_\a  \o_\a^2$ with its  average value $2e_0$. Finally, 
recalling eqs.~(\ref{calpha}) and approximating for large $N$ the sum over $\alpha$  with 
an integral ($\tau_\a\to\tau(q)$), we obtained \cite{noi}
\begin{widetext}
\begin{equation}
\frac{E(t)}{E(0)} = \frac{1}{\pi}  
                     \int_0^\pi e^{-2t/\tau(q)} \, dq =
                     e^{-t/\tau_{\scs 0}} 
                     I_{\scs 0} \left( \frac{\ds t}{\ds \tau_{\scs 0}} \right)
			   = \left\{
                        \begin{array}{ll}
                    e^{\ds -t / \tau_{\scs 0}}                         & 
                    \ {\rm for} \ t \ll \tau_{\scs 0},                   \\
                    \frac{\ds 1}{\ds \sqrt{2 \pi (t/\tau_{\scs 0}) }}   &
                    \ {\rm for} \ t \gg \tau_{\scs 0}.                                               
                        \end{array}
                    \right.
\label{Toten_free}
\end{equation}
\end{widetext}
where $I_{\scs 0}$ is the modified zero--order Bessel function, whose asymptotic expansions  have been
used to obtain the last equality. We conclude that relaxation is exponential only at short times 
while a crossover to a power--law decay occurs at $t \approx \tau_{\scs 0}$ . Moreover, despite the 
differences in the relaxation times (\ref{calpha}), the decay law for $E$ turns out to be the same in the
free and fixed--end systems.  

The above formulas must be taken with caution when dealing with a large but finite system. In such a
case, there is of course a lower cutoff at wavenumbers of order $\pi/N$. Therefore, after a time
of the order of the lifetime of the longest--lived mode,  $\tau_{\scs N-1}\sim N^3/\gamma$,
formula~(\ref{Toten_free}) no longer holds and a further crossover to the exponential decay law
$\exp(-2t/\tau_{\scs N-1})$ occurs. This has been also verified numerically~\cite{noi}.

\subsection{Non--linear lattices: the role of discreteness}

Since the first results of relaxation experiments in non--linear lattices were 
reported~\cite{Aubry1,Aubry2}, there has been some debate as to the nature of the decay law 
of the system energy and its relation with the spontaneously emerging localised modes. In
particular, it has been claimed that the phenomenon of energy pinning in the form of  discrete
breathers would result in a glassy--like relaxation, i.e. stretched exponential decay law. Here we
shall describe how the main role in determining the relaxation properties of  a non--linear
lattice is indeed played by the type of potential energy. In particular, we will highlight the
importance of the relative strength of the  inter--particle and on--site potentials,
i.e. the  degree of discreteness of the system.  In order to do so,  we study the
general class  of non--linear lattices obeying the equations of motion~(\ref{eq_mot}). In
particular, we shall report here the results of numerical simulations performed with
\begin{equation}
\label{UV}
V(x) = \frac12 x^2 + \frac14 x^4 \qquad U(x) = \frac14 \kappa x^4 \quad .
\end{equation}
In this case, the relative strength of local and inter--particle non--linearities
is accounted for by the parameter $\kappa$. 
\begin{figure*}[ht!]
\centering
\subfigure[]{
\includegraphics[width=7.5 truecm]{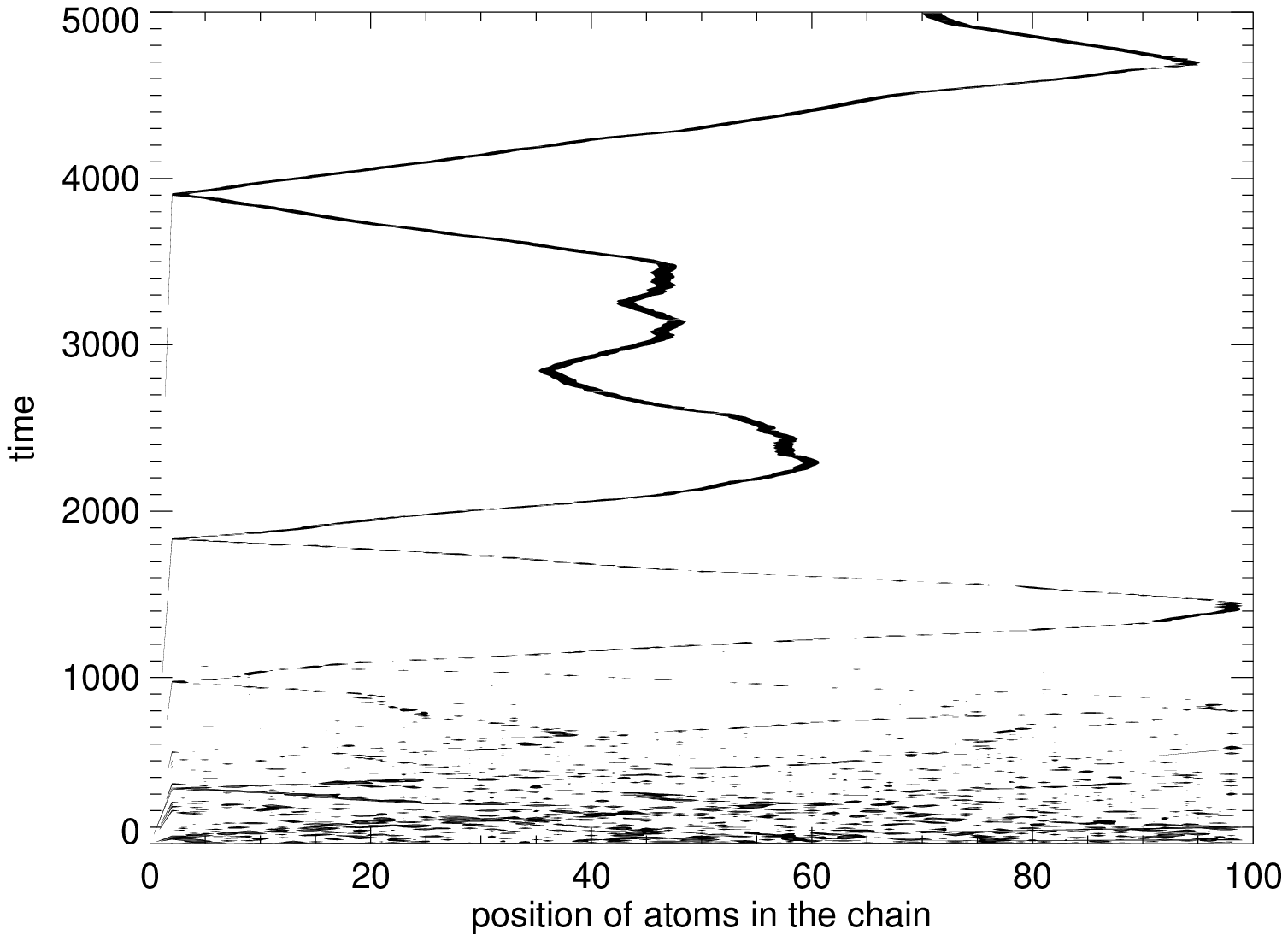}}
\subfigure[]{
\includegraphics[width=7.5 truecm]{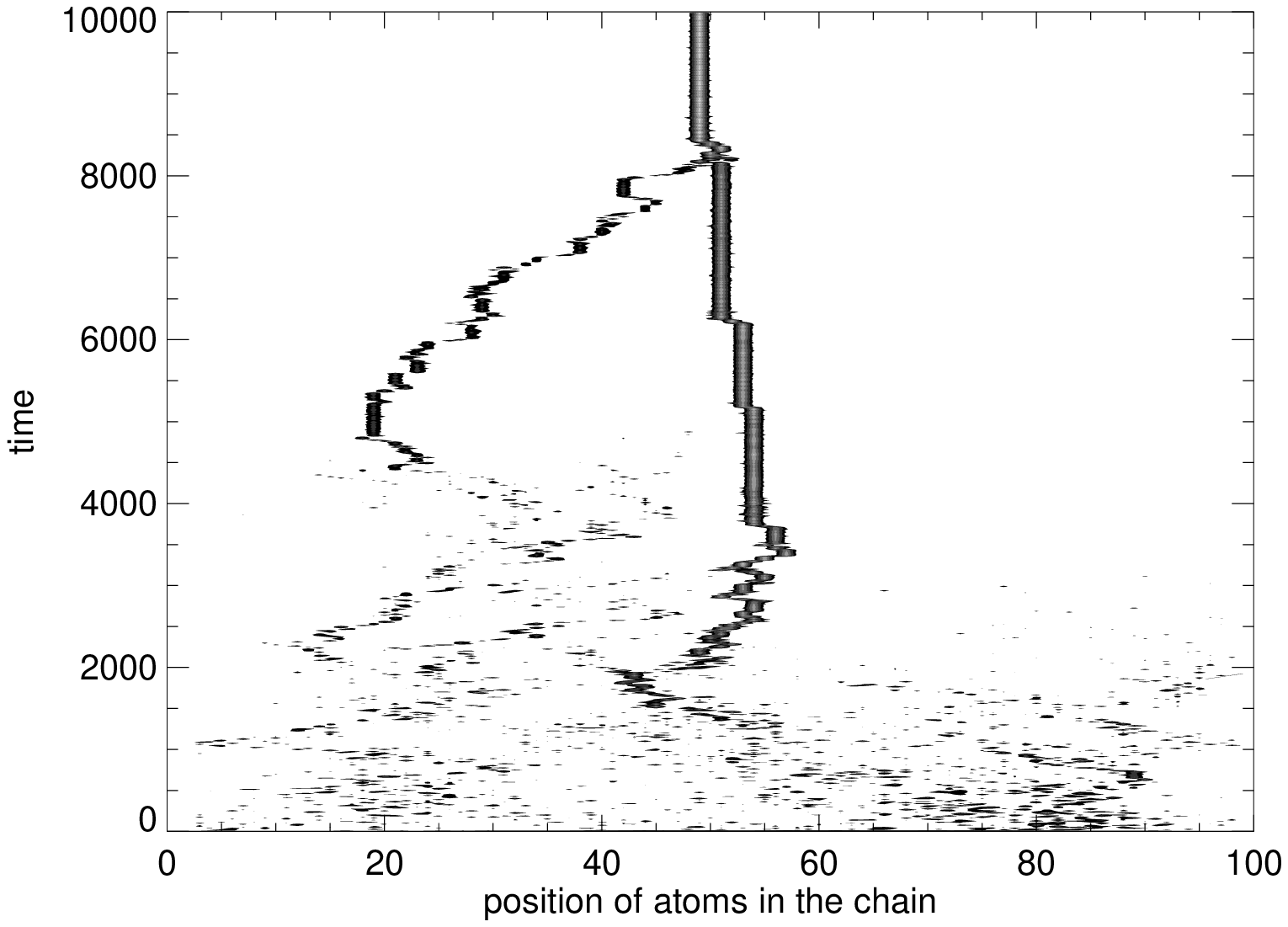}}
\caption{\small  \label{f:cpl} Relaxation in an FPU chain with 
quartic on--site potential, $\gamma=0.1$, $N=100$, $e_0=1$. 
Space--time contour plots of the symmetrised site energies. 
(a) $\kappa=0.1$. (b) $\kappa=10$}
\end{figure*}
\begin{figure*}
\centering
\subfigure[]{
\includegraphics[width=7.5 truecm, clip]{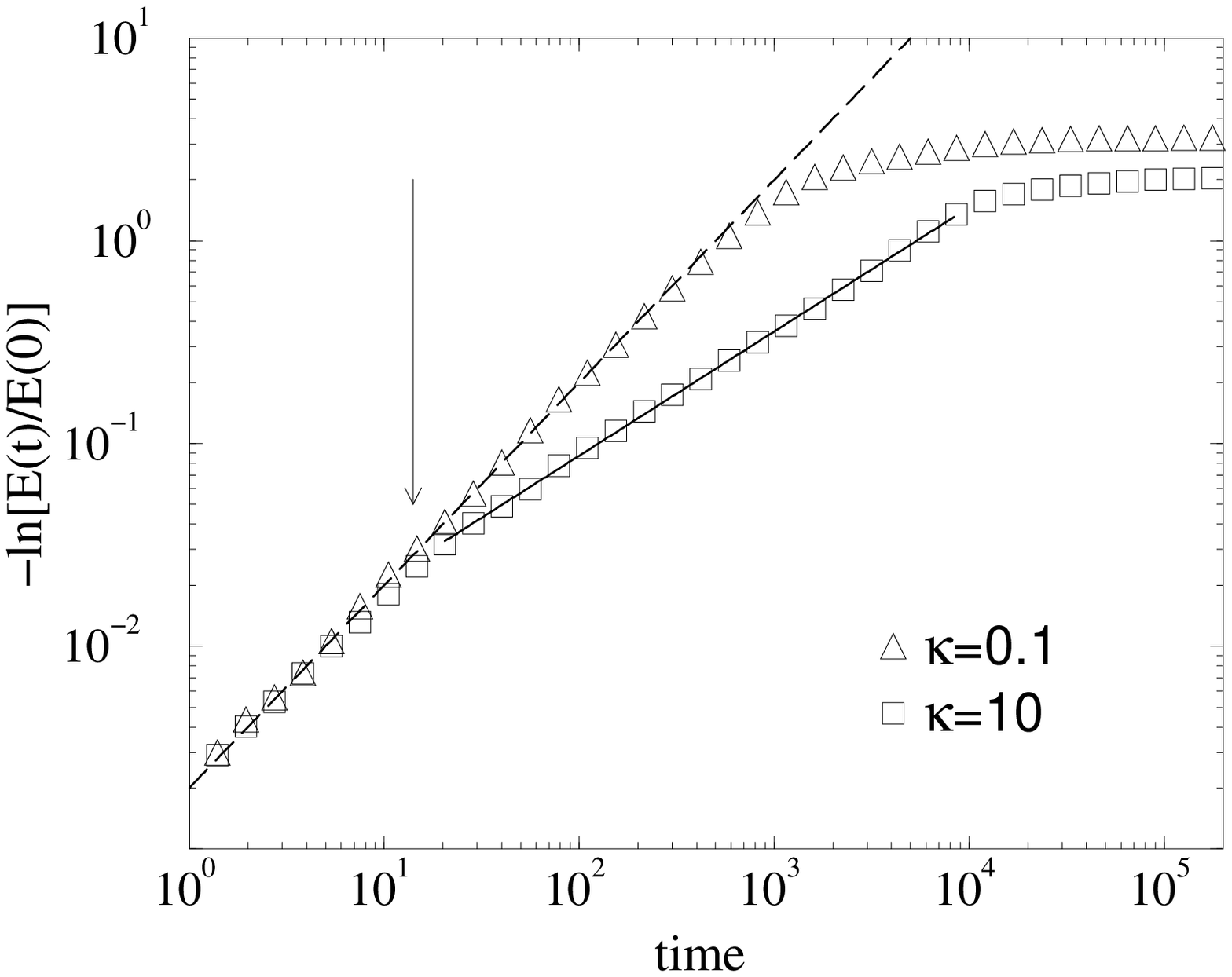}}
\hspace{0.5mm}
\subfigure[]{
\includegraphics[width=7.2 truecm]{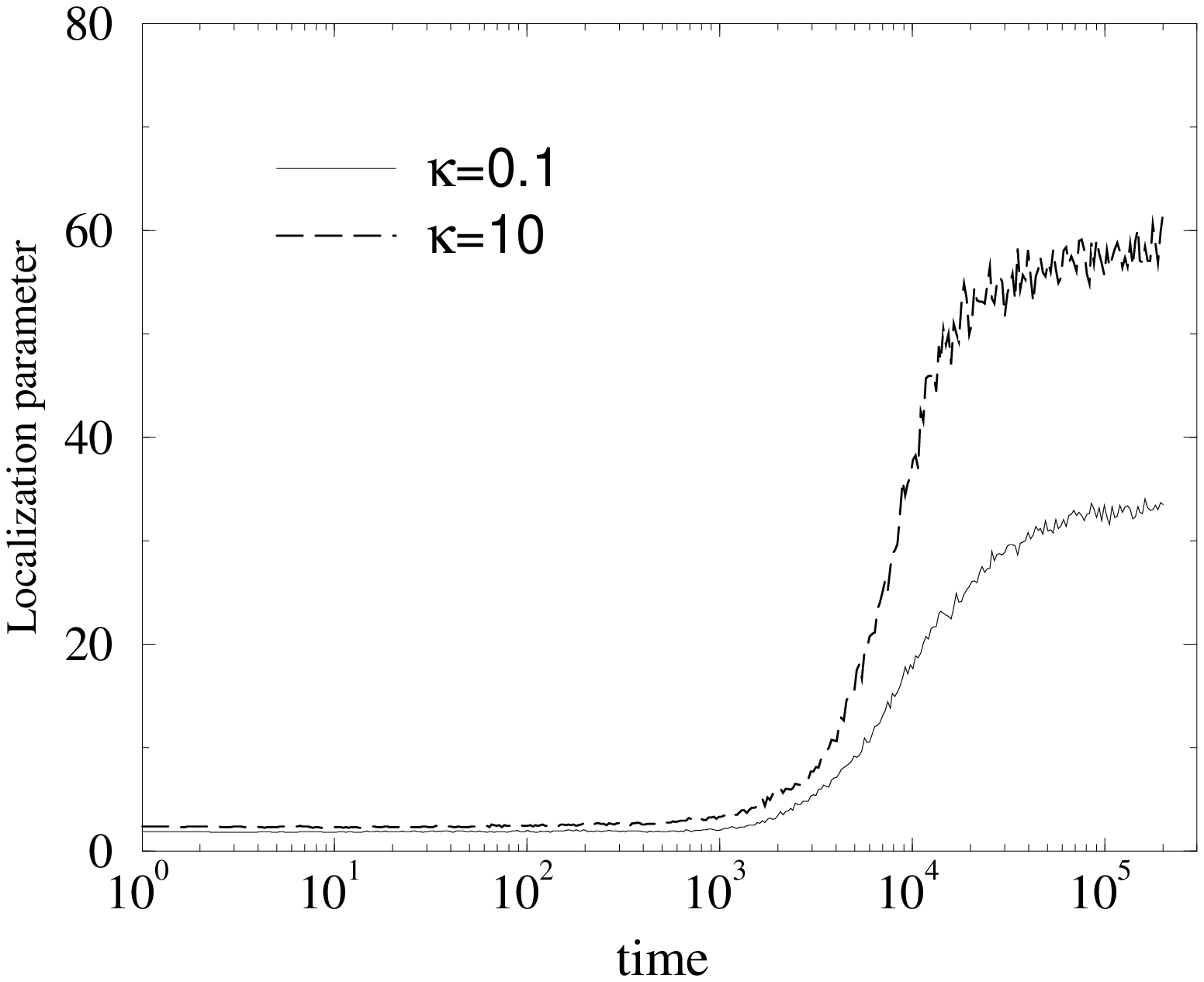}}
\caption{\label{osk_rel} \small FPU chain  
with quartic on--site potential, $N = 100$, $\gamma = 0.1$ and $e_0 = 1$.
(a) Plot of $\DLE$ vs time. The dashed line is the exponential $\exp(-t/\tau_{\scs 0})$, 
while the solid line 
is a fit with the stretched exponential law $\exp [-(t/\tau)^\sigma]$
$(\sigma \approx 0.61)$. The arrow marks the time at which the system with $\kappa = 10$ departs 
from the initial exponential trend.
(b) Localisation parameter vs time for different values of
the on--site coupling $\kappa$.
The data of both ${\cal D}$ and ${\cal L}$ are 
averaged over 32  initial conditions.}
\end{figure*}

In Fig.~\ref{f:cpl} we show the space--time contour plots of the symmetrised site energies, 
defined as
\[
h_p \; = \;\frac{1}{2} \dot{u}_p^2 + 
         \frac{1}{2} \left[ V(u_{p+1}-u_p) + V(u_p-u_{p-1}) \right] + U(u_p).
\]
The instantaneous total energy of the system is then given by
\(E = \sum_{p=0}^{N-1} h_p\).
We see that  a single localised excitation emerges from the relaxation process
for both small and large values of $\kappa$. The pathway to localisation
is the same as the one observed in the FPU model~\cite{noi}. 
In particular, the basic mechanism leading to localisation is modulational instability
of short wavelength modes. The latter can only be effective if dissipation of 
long wavelength phonons is fast enough. As it shows from Eq.~(\ref{calpha}), this occurs
only in the case of free--ends BC, whereas fixed--end BC strongly inhibit such process. 
On the other hand, breather mobility is strongly reduced in the  highly  discrete system (Fig.~\ref{f:cpl} (b)).
Such difference is even more evident in the decay law of the normalized energy 
$E(t)/E(0)$, which turns from pure exponential to stretched--exponential behaviour upon
increasing $\kappa$. This is better appreciated by plotting the indicator
\begin{equation}
\DLE (t) = - \ln [E(t)/E(0)] 
\label{DLEdef}
\end{equation}
in a log--log scale, so that 
a stretched--exponential law of the form 
$E(t)=E(0) \exp[-(t/\tau)^\sigma]$ becomes a straight line with slope $\sigma$ that intercepts 
the $y$--axis at $-\sigma \ln\tau$ (Fig.~\ref{osk_rel} (a)).
The scaling regions correspond to the onset of energy localisation, as can be seen 
by plotting the localisation parameter
\begin{equation}
{\cal L}(t) = N \, \frac{\ds \sum_p h_p^2(t)}
                     {\ds \left[ \sum_p h_p(t) \right]^2}.
\label{local}
\end{equation}
(Fig.~\ref{osk_rel} (b)).
From its very definition, the fewer sites the energy is localised onto, 
the closer ${\cal L}$ is to $N$ . On the other hand, the more evenly the energy is spread among all 
the particles, the closer ${\cal L}$ is to a constant of order 1.
It should be noted, that ${\cal L}$ is a relative quantity, and bears no information
on the amount of energy that is localised. As a matter of fact, the fraction of $e_0$
that gets trapped in the residual state is found to be greater the higher the degree of discreteness
of the system.

To summarise, if the on--site force is weak enough,
the behavior is basically the same as the FPU chain which, in turn, exhibits in the approach to the
residual state the same exponential
decay law of its linearised counterpart, i.e. $E(t)/E(0) \propto \exp
(-t/\tau_0)$ (see  Eqs.~(\ref{Toten_free})). The reasons for such behaviour are the same
discussed for the FPU model~\cite{noi}. First of all, an effective harmonic Hamiltonian with
energy--dependent renormalized frequencies is known to account for several equilibrium properties
of the FPU chain~\cite{Alabiso,Lepri}. Second, the harmonic approximation becomes increasingly
accurate as time elapses, simply because more and more energy is extracted from the system by the
reservoir. 
On the contrary, if the system is highly discrete, the first relaxation stage is indeed described
by a stretched exponential law. A quantitative explanation of it 
 is still lacking, but it is clear that in this case an effective
description of such genuine non--linear phenomenon in terms of quasi--harmonic modes breaks down. 
In other words, the resulting interactions among linear modes inhibit the energy flow
from the bulk to the boundaries, thus slowing down the dissipation.

The onset of a pseudo--stationary state corresponds in Fig.~\ref{osk_rel} to
the almost flat final portions of the curves. Actually, being the system
globally dissipative, energy is  still at that stage exponentially decreasing
but  with a huge time constant $\tau_b$. The latter is expected to be
inversely proportional to the breather amplitude at the lattice edges, which,
in turn, should be of order $\exp(-N/\ell)$, with $\ell\ll N$ being the
localisation length.

The above scenario is generic for large enough systems. Nonetheless, failure to spontaneously localise
energy occurs the more frequently the smaller is the lattice. This is because the probability of
occurrence of large enough energy fluctuations at a given energy density decreases with the number of
particles. In other words, given a number of different initial equilibrium conditions,  only a fraction
$n$ of them will give rise to breathers. To give a quantitative idea, for the FPU model with $e_0=1$,
$n$ is only about 50 \%  for a chain of $N=20$ particles but  rapidly approaches 100 \% already for $N
\gtrsim 100$. In the absence of spontaneous localisation, the non--linear systems show no difference with
respect to the linear ones and $E(t)$ decays according to Eqs.~(\ref{Toten_free}).  

\subsection{The effective--exponent analysis}

We have repeatedly mentioned that for the FPU model no stretched exponential relaxation
is observed, even in presence of a weak on--site force. In this section we wish to remark this
statement by reporting a more detailed analysis of the energy decay. This can be accomplished by
defining  an effective exponent $\zeta(t)$ through the logarithmic derivative
\begin{equation}
\label{beta}
\zeta (t) = \frac{d [ \ln \DLE (t)]}{d [\ln t]} \quad .
\end{equation}
It is clear from definition~(\ref{beta}) that if a portion of the energy decay curve goes as 
$E(t)=E(0) \exp[-(t/\tau)^\sigma]$ this would result in a plateau of height $\sigma$ in the 
corresponding $\zeta(t)$ curve. For the problem at hand, interpretation of numerical data with 
this indicator is however a delicate matter. Actually, we want to remark that a naive 
analysis can lead to misleading or even incorrect conclusions~\cite{reigada1,reigada2}.
 
The most convincing way to illustrate this is to discuss the case of the harmonic chain where 
localisation and stretched-exponential behaviour are obviously excluded a priori. From
Eq.~(\ref{Toten_free}), it is easily seen that $\zeta$ should start from 1 and monotonically vanish as
$1/\ln t$ for $t\gg \tau_0$. On the other hand, we already learnt that in a finite chain a
further crossover to an exponential law $\exp(-2t/\tau_{\scs N-1})$ occurs so that $\zeta$ must
approach again 1 for $t\gg \tau_{\scs N-1}$. The net results of those competing trends is that, for
a finite $N$, $\zeta$ displays a broad minimum $\zeta_{min}$, at $t \sim \tau_{\scs N-1}$ . The
value $\zeta_{min}$ (which is independent of $\gamma$) can be estimated by noting that from
Eqs.~(\ref{beta}) and~(\ref{Toten_free}) one has $\zeta(t) \approx 1/\ln[2 \pi t/\tau_0]$ for $t
\gg \tau_0$, and hence $\zeta_{min} \approx 1/2\ln[4N/\pi]$.  A simple plot of the curves shows how
the the minimum $\zeta_{min}$   can be incorrectly interpreted as a plateau if observed on a too
short time scale. Of course, the same scenario is likely to appear in relatively short FPU chains
whenever the chosen initial conditions do not yield localisation (see Fig.~\ref{f:beta}). 

\begin{figure*}[ht!]
\begin{center}
\subfigure[]{\includegraphics[width=8cm]{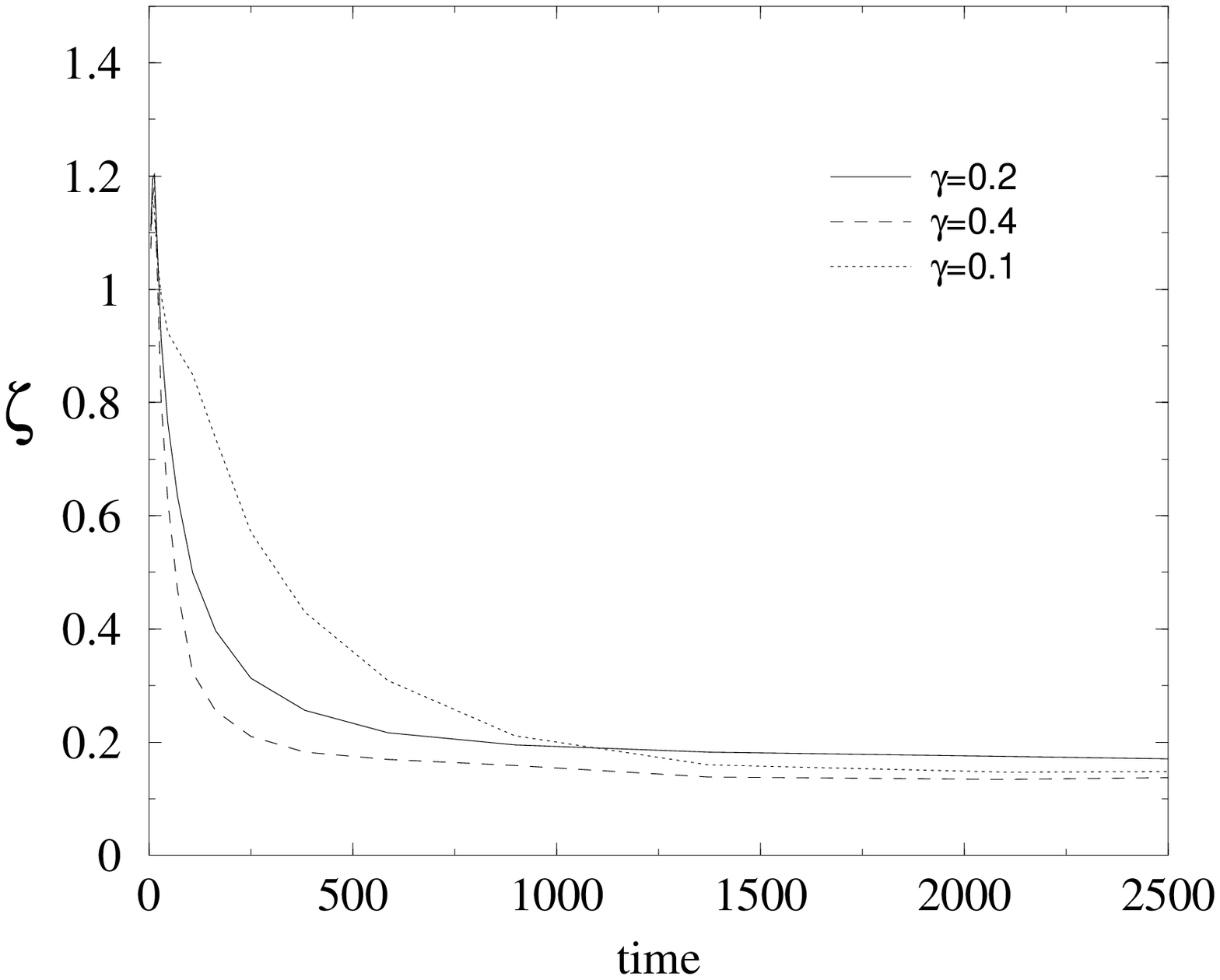}}
\subfigure[]{\includegraphics[width=7.65cm]{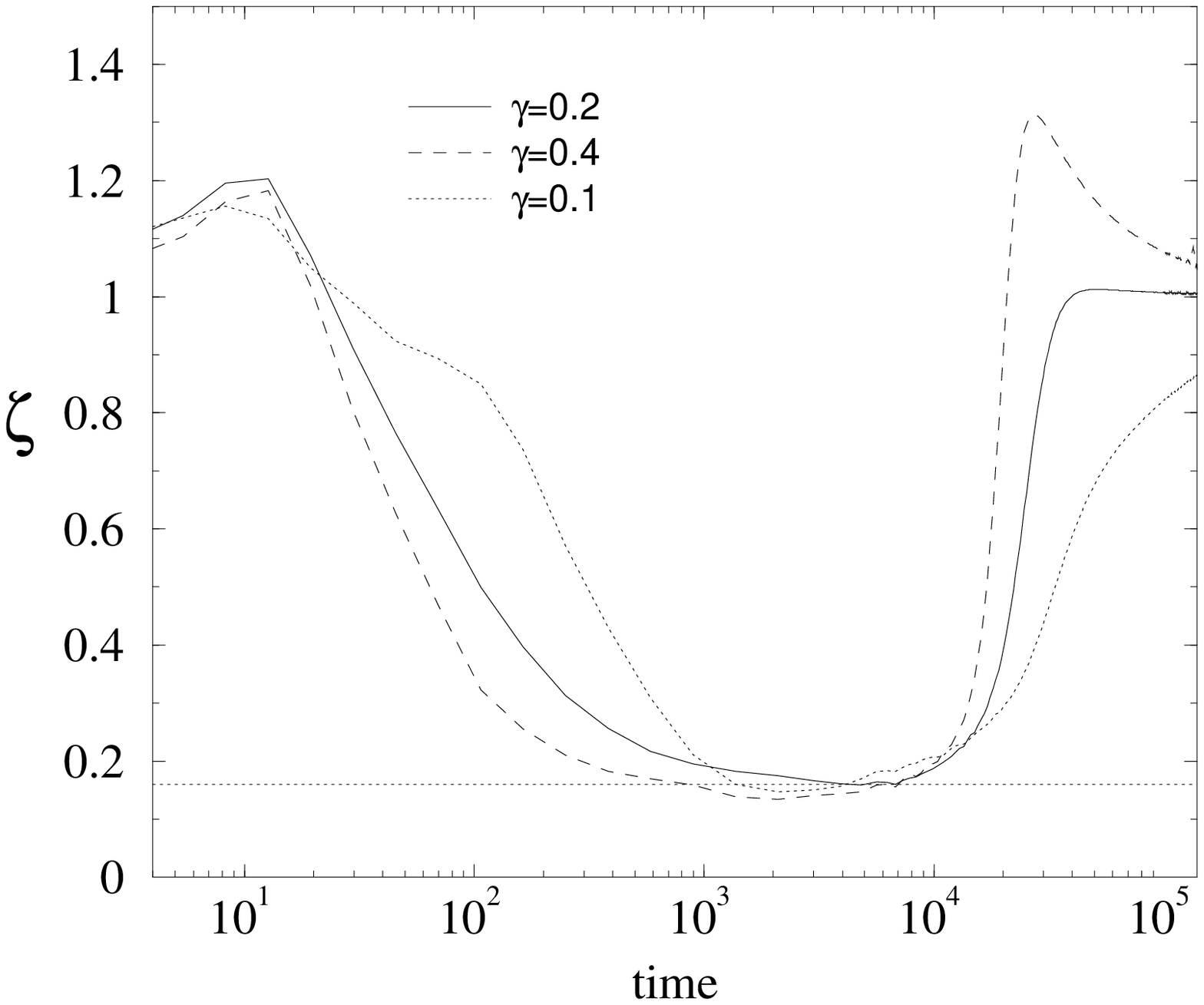}}
\caption{\small
Plot of $\zeta(t)$ for a short FPU chain ($N = 20$) for different $\gamma$s and $e_0 = 2.$.
(a) Short linear time scale. (b) Long logarithmic time scale. The dotted line
is the $\gamma$--independent minimum $\zeta_{min}$ (see text).
Localisation does not even occur for this initial condition but one can incorrectly conclude 
in favour of a stretched--exponential law.}
\label{f:beta}
\end{center}
\end{figure*}

When localisation does occur, the residual state decays exponentially~\cite{noi}. Therefore one expects
$\zeta$ to converge to 1 as the residual state is approached. Actually, one observes first very small
values of $\zeta$, while  such a convergence does occur only on much longer time scales. To see why,
let us consider a one--breather  residual state. The total energy decays as  $E(t) = E_b
e^{-t/\tau_b}$, where $E_b$ is the initial energy of the breather and $\tau_b$ its characteristic decay
time.  The time constant $\tau_b$ is huge:  in fact, it is roughly inversely proportional to the
breather amplitude at the edge sites of the chain. From definitions~(\ref{beta}) and~(\ref{DLEdef}) 
one has
\begin{equation} 
\label{beta_B} \zeta =
\frac{t/\tau_b}{t/\tau_b -\ln[E_b/E(0)]} \quad . 
\end{equation} 
Since $-\ln[E_b/E(0)]$ is a number of
order 1, we see that, at times such that  $t \ll \tau_b \ln[E_b/E(0)]$, $\zeta$ is very small.
The numerical results in this case are just a deceptive byproduct of the wrong normalization.  
The exponent $\zeta$ will start converging  to 1 only when $t \gg \tau_b \ln[E_b/E(0)]$, 
which is a huge time,
being also $E_b \ll E(0)$. In this case the $\zeta$ analysis may be misleading in
identifying the nature of the decay law. The way around such problem would be to calculate the
effective exponent by normalizing the total energy to $E_b$ . However, this solution is in turn
complicated by the intrinsic uncertainty in locating the time origin for the decay of the breather
state.

\section{\label{sec_2D} 2D lattices}

Since a careful study of the 2D Klein-Gordon model has already been presented~\cite{Aubry2}, we 
focus here on the 2D FPU model. In this respect, the most natural extension of model
(\ref{eq_mot}) should involve a two--component displacement vector. For the sake of simplicity,
we rather consider here a scalar model with only one degree of freedom $u_{i,j}$ per lattice site.
Actually, a preliminary series of simulations of the 2D vector version showed no appreciable
differences with what reported here, and will be discussed elsewhere. 

With the same choice of non--dimensional units introduced in section~\ref{sec_gen}, 
the model on a $N\times N$ lattice 
($i,j=0,\ldots,N-1$) with damping on all edges is defined by the equations of motion
\begin{equation}
\label{Ham2Dsc}
\begin{split}
\ddot{u}_{i,j} = \,  &V'(u_{i+1,j} - u_{i,j})-V'(u_{i,j} - u_{i-1,j}) + \nonumber\\
                     &V'(u_{i,j+1} - u_{i,j})-V'(u_{i,j} - u_{i,j-1}) - 
		 \sum_{p,q=0}^{N-1} \Gamma^{p,q}_{i,j} \dot{u}_{p,q}  ,
\end{split}	       
\end{equation}
where  $\Gamma^{p,q}_{i,j}= \gamma [g_{i,p}\d_{j,q}+\d_{i,p}g_{j,q}-g_{i,p}g_{j,q}]$,
with $g_{i,p}=\d_{i,p}[\d_{p,0} + \d_{p,N-1}]$.
Since localisation has been shown to be strongly inhibited by fixed--ends 
boundary conditions, we consider here free--ends BC. 

Calculations identical to the one described in section~\ref{sec_harm} can be easily extended
to the case of a simple 2D harmonic lattice. It turns out that the contributions from the two
spatial coordinates are identical for an $N \times N$ lattice, and therefore factorize, yielding
\begin{equation}
\label{Etot2D}
\frac{E(t)}{E(0)} = \left[
                     e^{-t /\tau_{\scs 0}}
                     I_{\scs 0} \left( \frac{\ds t}{\tau_{\scs 0}} \right)
                    \right]^2
		    \approx
		    \begin{cases}
		    e^{\ds -2t / \tau_{\scs 0}} & \text{for $t \ll \tau_{\scs 0}$} \\
		    \frac{\ds 1}{\ds 2 \pi (t/\tau_{\scs 0})} & \text{for $t \gg \tau_{\scs 0}$} 
		    \end{cases} \quad ,
\end{equation}
with $\tau_{\scs 0} = N / 2 \gamma$. In particular, the law~(\ref{Etot2D})
applies in the absence of localisation, as well as during the first stage before the onset of 
localisation.  In Fig.~\ref{f:en_2D} we plot the energy decay curves of an FPU lattice with $N=32$
relaxing from different values of the energy density $e_0$. 
The first stage of the decay indeed follows the theoretical prediction Eq.~(\ref{Etot2D}).
Moreover, it is clear that localisation
does not occur below a finite value of $e_0$ (located between $e_0=0.2$ and $e_0=0.3$ in this case).
In particular, in the absence of localisation the energy decay is again described by the 
two--crossover scenario. Notice that, also in this case, failure to spontaneously localise
energy results in a minimum of the effective exponent $\zeta$, due to the competition of 
the power law decay $1/t$ and the exponential law $\exp(-2t/\tau{\scs N-1})$.
\begin{figure}[ht!]
\centering
\includegraphics[width=8 truecm]{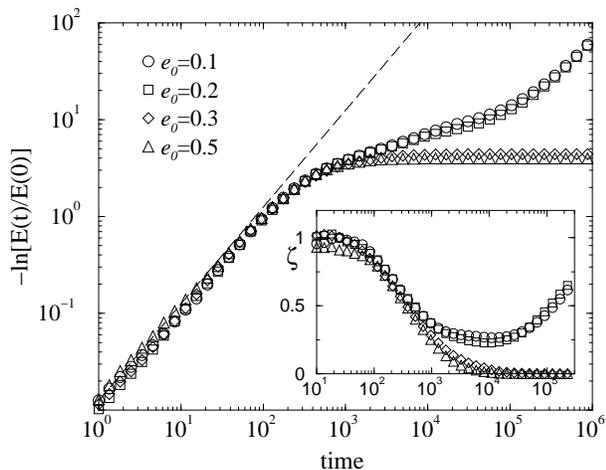}
\caption{\label{f:en_2D} 2D FPU lattice, $N=32$, $\gamma=0.1$. Plot of ${\cal D}(t)$ for
different values of the initial energy $e_0$ (symbols). The dashed line is 
a plot of Eq.~(\ref{Etot2D}) for $t \ll \tau_{\scs 0}$. The inset shows the corresponding 
effective exponents $\zeta(t)$ calculated from Eq.~(\ref{beta}). 
Notice the minimum associated with the absence of localisation.}
\end{figure}
\begin{figure}[hb!]
\begin{center}
\subfigure[]{
\includegraphics[width=8cm, height=9.25 truecm,clip]{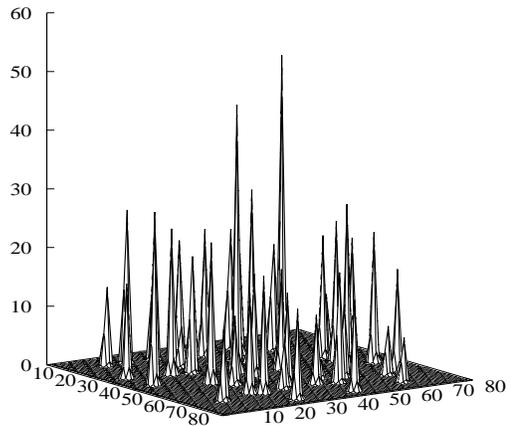}}
\subfigure[]{
\includegraphics[width=7.5cm]{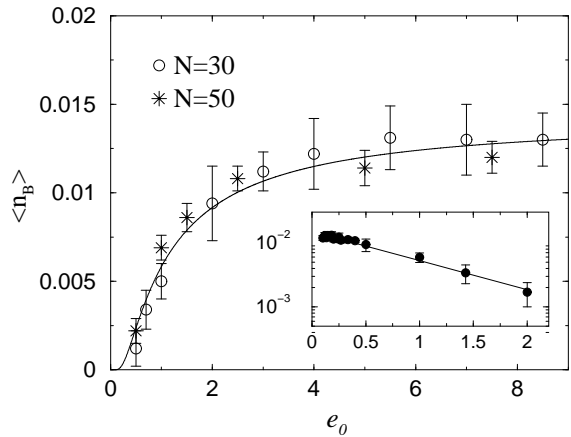}}
\caption{\small 
2D FPU lattice, $N=80$, $\gamma=0.1$. (a) $e_0=1$, 
symmetrised site energies in the residual state state.
(b) Average breather density 
$\langle n_{\scs B} \rangle$ vs initial energy density
for $N=30$ and $N=50$ and Arrhenius plot.
The inset shows the average density measured in the $N=50$ system 
vs $1/e_0$ in lin--log scale,
and an exponential fit.}
\label{f:2Dplot}
\end{center}
\end{figure} 
\subsection{Fluctuation--activated localisation}

The relaxation process described in terms of the energy decay curves is identical to the
two--crossover picture which applies to the 1D FPU case. Nonetheless, the scenario markedly differs
from the 1D case in two respects. First of all, the mobility of the localised excitations which
spontaneously emerge is drastically reduced. In fact, after a first stage in which strong
interaction is observed, they arrange on a ``random lattice'', which, on the time scale of our
typical simulations, appears indeed to be frozen. One of such states is illustrated in 
Fig.~\ref{f:2Dplot} (a). This is presumably due to a smaller ``scattering section'' in 2D. Obviously,
due to the presence of dissipation, this state decays on a much longer time scale, which (as
discussed above) scales exponentially with the linear size of the lattice.

The second important difference with the 1D case stems from the appearance of a finite energy
threshold $\Delta$ for the existence of breathers (akin to envelope solitons in the
small--amplitude limit) in 2D \cite{2D_tresh}. As a consequence, we expect that localised modes are
generated in  the relaxation dynamics only by fluctuations that are large enough to overcome such
threshold.  The spontaneous excitation of breathers can thus be seen as an activated process.
Accordingly, their number will be exponentially small in the ratio between $\Delta$ and some
quantity measuring the strength of fluctuations.  Therefore, one
expects the average density of breathers (i.e. the average number of breathers per lattice site) 
in the residual state to follow an Arrhenius law of the form
\begin{equation}
\label{arrhenius}
\langle n_{\scs B} \rangle \; \propto  \; \exp( -\beta \Delta) \quad .
\end{equation}
From Eq.~(\ref{arrhenius}), it is tempting to identify $\beta$ with some inverse
temperature. However, one has to keep in mind that we are dealing with a non--equilibrium process
and this identification can only make sense if energy release is adiabatically slow. If this is
true, $1/\beta$ should be proportional to (and smaller than) the initial  temperature,  which,
in turn, is roughly proportional to the initial energy density.
 
In our numerical simulations we clearly observe that $n_{\scs B}$ strongly depends on the initial
specific energy. In Fig.~\ref{f:2Dplot} (b) we plot  $\langle n_{\scs B} \rangle$ for two 
system sizes as a function of $e_0$,
along with a fit performed with the law $\langle n_{\scs B} \rangle = C \, \exp(-\Delta^{\pr}/e_0)$, with
$\Delta^{\pr} \propto \Delta$. As it shows, the agreement is good. We
conclude that the spontaneous localisation of energy in our system is indeed  an 
activated process. In particular, the fit gives $\Delta^{\pr} \approx 0.9$. We note that a quantitative
analysis of the time $t_0$ required to reach the residual state (localisation time) reveals that 
$t_0 \propto 1/\langle n_{\scs B} \rangle \propto \exp(\beta  \Delta)$~\cite{Esc_proc}. By fitting
the numerical data for the localisation time, it is possible to get an independent confirmation of
the thermal activation scenario. In particular, we get $\Delta^{\pr} \approx 0.7$, in good agreement
with the value obtained above.
\subsection{Statistics of breathers' energies}

It is important to realize that the ``pseudo--stationary'' distribution of breathers 
in the residual state
emerges as a result of two competing mechanisms, namely the birth out of fluctuations
from the homogeneous state and the decay due to both the coupling with the external 
reservoir and to the various inelastic interactions (phonon--breather,
breather--breather etc). Although the former can be safely neglected, it is 
very difficult to understand the latter in full detail. Indeed, to the best of our 
knowledge, only a few studies are available~\cite{Johansson,Bang} and a full description
in terms of elementary interactions seems hardly feasible.
Nonetheless, simple statistical arguments can be of help in understanding the energy 
distribution to a greater detail. Let us consider the fraction of breathers 
${\cal P}(\e,t)$ having energy between $\e$ and $\e+d \e$ with $\e \ge \Delta$
(in the FPU model breathers of arbitrarily large energy exist due to the unboundedness
of the interaction potential)  
and let us define 
$B(\e)$ and  $D(\e)$ to be some phenomenological birth and 
decay rates, respectively. Furthermore, let us assume that birth of a breather can only 
occur by a spontaneous fluctuation from the ground state ($\e=0$). 
Stationarity requires the flux--balance condition
\begin{equation} 
B(\e){\cal P}(0) \;=\; D(\e){\cal P}(\e)
\end{equation}
Since the birth is an activated process we expect $B(\e) \propto \exp[-\beta \e ]$~\footnote{In
principle, one could have  a further energy--dependent prefactor in front of the exponential. That would
affect the form of the distribution. We neglect it for simplicity. }. This assumption yields (for a constant ${\cal P}(0)$)
\begin{equation} 
 {\cal P}(\e) \;=\;  \frac{{\cal P}(0)}{D(\e)} \,\exp(-\beta \e )
 \label{prob} 
\end{equation}
In this simplified description, the prefactor of the exponential term  is thus interpreted
as a measure of the effective breather lifetime $\tau(\e)$ (${\cal D}(\e) \propto 1/\tau(\e)$)
and is a priori unknown. In all simulations we
observe that small--amplitude breathers decay more easily. This observation agrees
with the accepted existence of a preferred energy flow from small--amplitude
breathers to large--amplitude ones in breather--breather collisions~\cite{Bang}.
Moreover, Eq.~(\ref{prob}) requires that $\tau(\Delta)=0$.
Thus, we postulate 
\begin{equation} 
  D(\e) \;\propto\; (\e - \Delta)^{-z} \quad ,
\label{espz} 
\end{equation}
where the exponent $z>0$ can be estimated by measuring the distribution of breather
energies in the vicinity of the threshold $\Delta$.
We can estimate the average density of breathers in the residual state from
Eqs.~(\ref{prob}) and~(\ref{espz}) as
\begin{equation} 
\langle n_{\scs B} \rangle \;=\; \int_\Delta^\infty {\cal P}(\e) \, d \e  
                                 \propto  \beta^{-(z+1)} \, e^{-\beta \Delta} .
\end{equation} 
This result is consistent with our initial hypothesis (Eq.~(\ref{arrhenius})).
 
From Eqs.~(\ref{prob}) and~(\ref{espz}) it follows that the average breather energy $\langle \e \rangle$
can be expressed as a function of $\beta$ and $\Delta$. Hence, we can write the  normalised  distribution
Eq.~(\ref{prob}) as a function of the three parameters $z$, $\langle \e \rangle$ and $\Delta$ in the
following fashion 
\begin{equation} 
\label{prob_norm} 
{\cal P}(\e) = \frac{(z + 1)^{z+1}} {\Gamma(z+1)
(\langle \e \rangle-\Delta)} \,  \left[ \frac{\e - \Delta}{\langle \e \rangle-\Delta} \right]^z \,  
e^{-(z+1) \left[ \frac{\ds \e - \Delta}{\ds \langle \e \rangle - \Delta} \right]} \quad , 
\end{equation} 
where $\Gamma(z)$ is the gamma function and $\langle \e \rangle = \Delta + (z+1)/\beta$. From the point of view
of the numerics, it is more accurate to deal with the cumulative (integrated) distributions. These are
easily obtained from the histograms without actually performing the integration, by noting that they are
nothing but rank--size plots with the axes inverted. The cumulative distribution of the
function~(\ref{prob_norm}) is 
\begin{equation} 
\begin{split}
\label{prob_cum} {\cal C}(\e) = \int_\e^\infty  {\cal P}(\e^{\pr}) d \e^{\pr} \;=\;  
                                    &\frac{1}{\Gamma(z+1)}  \\
                                    &\times \gamma \left(
z+1,(z+1)\left[  \frac{\e - \Delta}{\langle \e \rangle -\Delta}  \right] \right) 
\end{split}
\end{equation} 
where $\gamma(z,\e)$ is the incomplete gamma function, defined as  
\begin{equation} 
\label{gamma} 
\gamma(z,\e) =
\int_\e^\infty y^{z-1} e^{-y} dy \quad . 
\end{equation}
In principle one could evaluate $\Delta$ explicitly by calculating  the energy of exact breather solutions of
vanishing amplitude~\cite{2D_tresh}.  However, we expect that $\Delta \ll \langle \e \rangle$. Therefore, if the
population of breathers used to calculate the experimental distribution ${\cal C}(\e)$ is large enough, we
can assume that the smallest breather energy recorded $\e_{min}$ is close to the threshold $\Delta$.
Hence, we can perform a two--parameter fit with the theoretical prediction Eq.~(\ref{prob_cum}) with
$\Delta=\e_{min}$. 
\begin{figure}[ht!]
\centering 
\includegraphics[width=8.5truecm,clip]{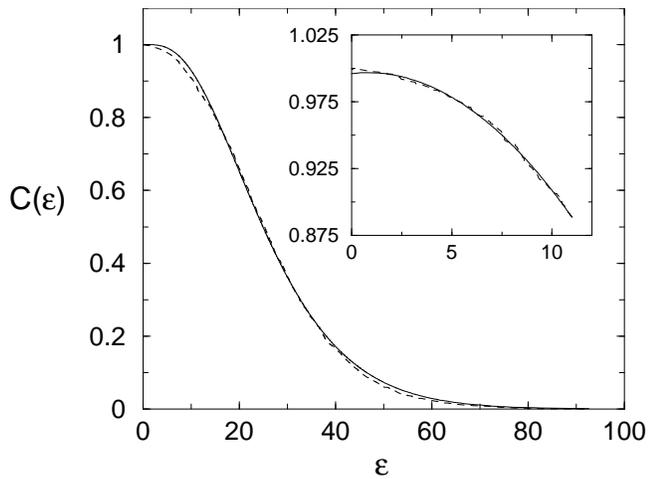} 
\caption{\small  2D FPU lattice,
$N=70$, $\gamma=0.1$ and $e_0=1$.  Experimental cumulative distribution of breather energies (dashed line)
and fit with formula~(\ref{prob_cum}) (solid line).  Both curves have been shifted to  the left
by $\e_{min}$. The breather population is here of 1742 breather
``events'', recorded in 50 different realisations of the initial condition. The inset shows a quadratic
fit of the small--energy portion of the distribution.} 
\label{f:N70pcum} 
\end{figure} 
The quality of such fits is illustrated in Fig.~\ref{f:N70pcum} for a lattice with
$N=70$.  As it shows, the agreement of our simple model with the numerics is excellent. In particular, the
analysis of the small--energy portion of the distributions yields $z=2$ (see inset in
Fig.~\ref{f:N70pcum}). We stress that the result of our fits were well reproduced irrespective of the particular
realisation of
$\e_{min}$. The fit of the experimental distributions  over the whole range of available energies
yields a slightly greater value, $z \approx 2.6$. This is because the large--energy tail of
the distributions accounts for the rarest events, which are under--represented within reasonably large
breather populations~\footnote{The main constraint comes from the time required to numerically integrate
the equations of motion of large lattices for a large number of realisations of a given initial
condition.}. In order to convince oneself that this is the case, it is enough to  introduce an
energy--dependent weight function in the fit, which gauges the relative weights of the small--energy and
large--energy portions of the curves.  In this case, the best estimates of $z$ monotonically decrease
towards the value $z=2$ upon lowering the relative weight assigned to the large--energy region.
\begin{figure}[ht!]
\centering
\includegraphics[width=8cm]{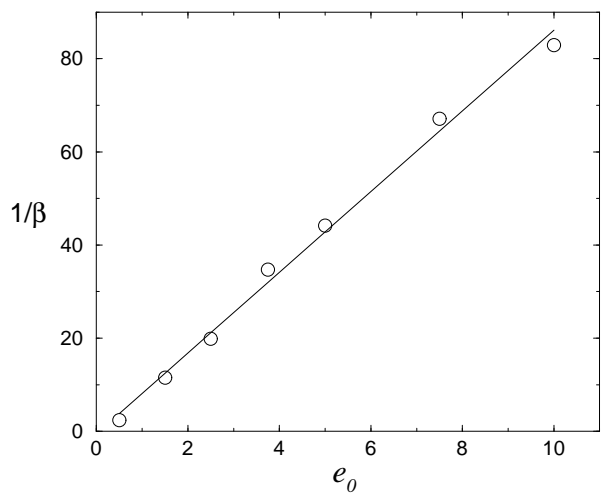}
\caption{\small 2D FPU lattice, $\gamma=0.1$, $N=50$.
Plot of Eq.~(\ref{beta_exp}) evaluated from the best--fit
values of the floating parameters $z$ and $\langle \e \rangle$ vs initial energy density
(symbols) and linear fit.}
\label{f:betavsE0}
\end{figure} 

The analysis of energy distributions also provides an independent confirmation that localisation
is in the present case an activated process.  To check this, we fitted the
distributions obtained by letting systems of given size relax from different values of $e_0$.
Then,  we evaluated $\beta$ from the best--fit values of the fitting parameters $z$
and $\langle \e \rangle$ according to the relation
\begin{equation}
\label{beta_exp}
 \frac{1}{\beta} = \frac{\langle \e \rangle - \e_{min}}{z+1} \quad .
\end{equation}
As shown in Fig.~\ref{f:betavsE0},  $\beta$ turns out to be  inversely
proportional to $e_0$ as expected.

\section{Conclusions}

The mechanisms yielding the spontaneous formation of 
localised periodic excitations, i.e. breathers, in 
spatially discrete non--linear dynamical systems are of primary importance
for the understanding of their physical interest.
In this paper we have focused on the phenomenon of 
relaxation to breather states by cooling from the boundaries
1D and 2D lattices. We have provided a detailed description
of the many facets of this phenomenon by combining numerical
studies with theoretical arguments.
In particular, we have shown that in models with sufficiently weak
``substrate'' forces the energy loss process
can be explained on the basis of a simple perturbative
analysis, that applies independently of the boundary conditions
and of the nature of the nonlinearity. In this sense, we can
claim that this is a general feature of this class
of models, where an initial exponential decay is followed by
a power--law behaviour. 
In the second stage of the decay process, boundary conditions play a crucial role
in allowing for the formation of a long--living breather state, when they
are taken free. On the contrary, fixed ends yield complete cooling
of the lattice eventually ruled by the exponential decay rate 
of the longest--lived Fourier mode.
We have also pointed out that the presence of a sufficiently
strong ``substrate'' force can turn the initial exponential decay
to a stretched--exponential law, while maintaining all the other
features of the previously described scenario. A theoretical
explanation of the origin of this
stretched exponential behaviour and its dependence on the strength
and on the nature of the local potential is a problem 
that deserves further and more refined investigations.
We have also commented about the technical difficulties that can
be encountered for identifying the correct time behaviour in
these models, where the crossover between different regimes can
be easily mistaken for an indication in favour of other scaling laws.

Another crucial aspect that we have widely analysed concerns the
main differences between 1D and 2D systems. At variance with
the former case, in 2D we have found numerical evidence
of the existence of a finite energy threshold for breather formation.
We have also proposed a statistical model that provides an effective
agreement with numerical results concerning the breather residual
state. This state eventually sets in as 
an almost static stationary configuration, where breathers of
different amplitudes coexist. The statistical model is based
on the hypothesis that this state emerges as a result of the
competition between breather activation and mutual interaction.
Moreover, it allows to determine an effective breather 
lifetime and its dependence on energy.     
In this framework the initial energy density plays the role
of the control parameter regulating the strength of fluctuations.

In our opinion these results provide a satisfactory understanding of
the phenomenon of spontaneous breather formation by cooling.
As a final remark, it seems worth investigating experimentally
the capability of real systems to store energy 
in the form of long--lived localised excitations.

\begin{acknowledgments}
This work has been supported by the European Union under the RTN project
LOCNET, Contract No. HPRN-CT-1999-00163. Francesco Piazza wishes to acknowledge 
numerous useful discussions with Katja Lindenberg.                               
\end{acknowledgments}

\end{document}